\documentstyle[epsfig,12pt]{article}

\begin{document}
\vskip 0.2in
\begin{center}
{\Large {\bf Correlation femtoscopy of multiparticle
processes}}\footnote{
\uppercase{W}ork
supported by grant 202/01/0779 of the \uppercase{G}rant
\uppercase{A}gency of the \uppercase{C}zech \uppercase{R}epublic.}
\end{center}
\vskip 0.15in
\begin{center}
R.~Lednick\'{y}\\
{\it Joint Institute for Nuclear Research,
Dubna, Moscow Region, 141980, Russia\\
Institute of Physics,
Na Slovance 2, 18221 Prague 8, Czech Republic}\\
\end{center}

\begin{abstract}
Recent results on particle momentum and spin correlations
are discussed in view of the role played by the
effects of quantum statistics, including
multiboson and coherence phenomena, and
final state interaction.
Particularly, it is demonstrated that the latter allows for
(i) correlation femtoscopy with unlike particles;
(ii) study of the relative space--time asymmetries in the
production of different particle species ({\it e.g.}, relative
time delays or spatial shifts due to collective flows);
(iii) study of the particle strong interaction hardly
accessible by other means ({\it e.g.}, in $\Lambda\Lambda$ system).
\end{abstract}

\section{Introduction}

The momentum correlations of particles at small relative velocities
are widely used to study space-time characteristics of the
production processes, so serving as a correlation femtoscope.
Particularly, for non-interacting identical particles, like photons
or, to some extent, pions, these correlations result from the
interference of the production amplitudes due to the
symmetrization requirement of quantum statistics
(QS) \cite{GGLP60,KP72}.
There exists \cite{KP75} a deep analogy of the momentum QS
correlations of photons with the space--time correlations
of the intensities of classical electromagnetic fields used in
astronomy to measure the angular radii of stellar objects
based on the superposition principle - so called HBT
intensity interferometry \cite{hbt}.\footnote
{
This analogy is sometimes misunderstood and the momentum
correlations are mixed up with the space--time HBT correlations
in spite of their orthogonal character and the failure of the
superposition principle for correlations of identical
fermions.
In fact, in spite of the common QS origin
of the momentum correlations of identical particles
and the space--time HBT correlations
(allowing for a generalization of the latter to any type
of identical bosons or fermions),
the corresponding correlation measurements differ in principle
\cite{KP75}
(see also \cite{lyu98}).
The former, being the momentum--energy measurement,
yields the space--time picture of the source, while the latter
does the opposite.
In particular, the dependence of the number of coincident
two--photon counts on the distance between detectors
(a quantum analogy of the HBT measurement)
provides the information on the
characteristic relative three-momenta of emitted photons and so,
when divided by the mean detected momentum, on the angular
size of a star but, of course, -
no information on the star radius or lifetime.
}

The momentum QS correlations were first
observed as an enhanced production of the
pairs of identical pions with small opening angles
(GGLP effect \cite{GGLP60}). Later on,
Kopylov and Podgoretsky \cite{KP72}
settled the basics of correlation femtoscopy; particularly, they
suggested to study the interference effect in terms of the
correlation function and clarified the role of the space--time
characteristics of particle production in various physical situations.

The momentum correlations of particles emitted at nuclear distances
are also influenced by the effect of
final state interaction (FSI) \cite{koo,gkw79,ll1}.
Thus the effect of the Coulomb interaction dominates the correlations
of charged particles at very small relative momenta
(of the order of the inverse Bohr radius of the two-particle system),
respectively suppressing or
enhancing the production of particles with like or unlike charges.
Though the FSI effect complicates the correlation analysis,
it is an important source of information allowing for the
coalescence femtoscopy
(see, {\it e.g.}, \cite{sy81,lyu88,mro93,sh99}),
the correlation femtoscopy with unlike particles \cite{ll1,BS86}
including the access to the relative space--time
asymmetries in particle production \cite{LLEN95} and a study
of particle interaction hardly accessible by other means.

We do not touch here the fluctuation measures which are closely
related with particle correlations in momentum space and
carry an important information on the dynamics
and space-time evolution of the production process
(see \cite{jk03} for a recent review).

The rest of the report is organized as follows.
In section 2, we briefly review the formalism of
particle correlations at small relative velocities.
The basic concepts of femtoscopy with identical
and nonidentical particles, including the access to the
relative space-time shifts in the emission of various
particle species,
and some recent results are reviewed in sections 3, 5 and 7.
In section 4, we discuss the present theoretical and
experimental status of the multiboson and coherence
phenomena in multiparticle production.
Recent results from correlation measurements of the
strong interaction in various two-particle systems
are reviewed in section 6.
In section 8, we briefly discuss spin correlations as a new
femtoscopy tool.
We conclude in Section 9.

\section{Formalism}
The ideal two-particle
correlation function ${\mathcal R}(p_{1},p_{2})$
is defined as a
ratio of the measured two-particle distribution to the
reference one which would be observed in the absence of the
effects of QS and FSI.
In practice, the reference distribution is usually constructed
by mixing the particles from different events of a given class,
normalizing the correlation function to unity at sufficiently
large relative velocities.

Usually, it is assumed
that the correlation of two particles
emitted with a small relative velocity
is influenced by the effects of
their mutual QS and FSI only\footnote
{Besides the events with a large phase-space density fluctuations,
this assumption may not be justified also in low energy
heavy ion reactions
when the particles are produced in a strong Coulomb field of residual
nuclei. To deal with this field a quantum adiabatic (factorisation)
approach can be used \cite{3body}.
}
and that the momentum dependence of the one-particle emission
probabilities
is inessential when varying the particle four-momenta
$p_{1}$ and $p_{2}$ by the
amount characteristic for the correlation due to QS and FSI
({\it smoothness assumption}). Clearly, the latter assumption,
requiring the components of the mean space-time distance
between particle emitters
much larger than those of the space-time extent
of the emitters, is well justified for heavy ion collisions.

The correlation function is then given by a square of the properly
symmetrized Bethe-Salpeter
amplitude in the continuous spectrum of the two-particle states,
averaged over the four-coordinates $x_{i}=\{t_{i},{\bf r}_{i}\}$
of the emitters and over the total spin $S$ of the two--particle
system \cite{ll1}.
After the separation of the unimportant phase factor
due to the c.m.s. motion,
this amplitude reduces to the one depending
only on the relative four-coordinate
$\Delta x\equiv x_1-x_2=\{t,{\bf r}\}$
and the generalized relative momentum
$\widetilde{q}=q-P(qP)/P^2$, where $q = p_1-p_2$,
$P=p_1+p_2$ and $qP = m_1{}^2-m_2{}^2$;
in the two-particle c.m.s., ${\bf P} = 0$,
$\tilde q = \{0,2{\bf k}^*\}$ and $\Delta x = \{t^*,{\bf r}^*\}$.
At equal emission times of the two particles in their c.m.s.
($t^* \equiv t_1^*-t_2^* = 0$),
the reduced non--symmetrized amplitude coincides with
a stationary solution $\psi ^{S(+)}_{-{\bf k}^*}({\bf r}^*)$ of the
scattering problem having at large distances
${r}^*$ the asymptotic form of a
superposition of the plane and outgoing spherical waves
(the minus sign of the vector ${\bf k}^{*}$ corresponds to the reverse
in time direction of the emission process).
The Bethe-Salpeter amplitude can be usually substituted  by
this solution ({\it equal time} approximation)\footnote
{
For non--interacting particles, the non--symmetrized Bethe-Salpeter
amplitude reduces to the plane wave
${\rm e}^{i\tilde qx/2}\equiv {\rm e}^{-i{\bf k}^*{\bf r}^*}$
which is independent of the relative time in the two--particle c.m.s.
and so, coincides with the corresponding equal--time amplitude.
For interacting particles, the {\it equal time} approximation is
valid on condition \cite{ll1}
$ |t^*|\ll m_{2,1}r^{*2}$ for
${\rm sign}(t^*)=\pm 1$ respectively.
This condition is usually satisfied
for heavy particles like kaons or
nucleons. But even for pions, the $t^{*}=0$ approximation
merely leads to a slight overestimation (typically $<5\%$) of the
strong FSI effect and,
it doesn't influence the leading zero--distance
($r^{*}\ll |a|$) effect of the Coulomb FSI.
}
so, for non--identical particles,
\begin{equation}
{\cal R}(p_{1},p_{2})\doteq
\sum_{S}\widetilde{\rho}_{S}
\langle |\psi_{-{\bf k}^{*}}^{S(+)}({\bf r}^{*})|^{2}
\rangle _{S};
\label{1}
\end{equation}
for identical particles, the amplitude in Eq.~(\ref{1})
enters in a symmetrized form:
\begin{equation}
\label{sym}
\psi_{-{\bf k}^{*}}^{S(+)}({\bf r}^{*}) \rightarrow
[\psi_{-{\bf k}^{*}}^{S(+)}({\bf r}^{*})+(-1)^{S}
\psi_{{\bf k}^{*}}^{S(+)}({\bf r}^{*})]/\sqrt{2}.
\end{equation}
The averaging in Eq.~(\ref{1})
is done over the four--coordinates of the emitters
at a given total spin $S$ of the two--particles,
$\widetilde{\rho}_{S}$ is the corresponding population probability,
$\sum_{S}\widetilde{\rho}_{S} = 1$.
For unpolarized particles
with spins $s_{1}$ and $s_{2}$ the probability
$\widetilde{\rho}_{S}=(2S+1)/[(2s_{1}+1)(2s_{2}+1)]$.
Generally, the correlation function is sensitive to particle
polarization. For example, if two spin-1/2 particles are
initially emitted with
polarizations $\mbox{\boldmath${\cal P}$}_1$ and
$\mbox{\boldmath${\cal P}$}_2$ then \cite{ll1}
\begin{equation}
\label{pol}
\widetilde{\rho}_0=(1-\mbox{\boldmath${\cal P}$}_1\cdot
\mbox{\boldmath${\cal P}$}_2)/4,~~~ \widetilde{\rho}_1=(3+
\mbox{\boldmath${\cal P}$}_1\cdot\mbox{\boldmath${\cal P}$}_2)/4.
\end{equation}

\section{Femtoscopy with identical particles}
For identical pions or kaons, the effect of the strong FSI is usually
small and the effect of the Coulomb FSI can be in first approximation
simply corrected for (see \cite{slape} and references therein).
The corrected correlation function is determined by the
QS symmetrization only (see Eq.~(\ref{sym}) and substitute the
non--symmetrized amplitude by the plane wave
${\rm e}^{i qx/2}$):
\begin{equation}
{\cal R}(p_{1},p_{2})= 1+\langle \cos(q\Delta x)\rangle .
\label{qscf}
\end{equation}
Its characteristic feature is
the presence of the interference maximum at small
components of the relative four-momentum $q$
with the width reflecting the inverse space-time extent
of the effective production region.
For example, assuming that for a
fraction $\lambda$ of the pairs,
the pions are emitted independently according to
one--particle amplitudes of a Gaussian form characterized by the
space--time dispersions $r_0^2$ and $\tau_0^2$
while, for the remaining fraction $(1-\lambda)$
related to very long--lived sources ($\eta$, $\eta'$, $K^0_s$,
$\Lambda$, \dots), the relative distances $r^*$ between the emitters
in the pair c.m.s. are extremely large,
one has
\begin{equation}
{\cal R}(p_{1},p_{2})=
1+\lambda\exp\left(-r_0^2{\bf q}^2-\tau_0^2q_0^2\right)=
1+\lambda\exp\left(-r_0^2{\bf q}_T^2-(r_0^2+v^2\tau_0^2)q_L^2\right),
\label{5}
\end{equation}
where $q_T$ and $q_L$ are the transverse and longitudinal
components of the three--momentum difference ${\bf q}$
with respect to the direction
of the pair velocity ${\bf v}={\bf P}/P_0$.
One may see that, due to the on-shell constraint \cite{KP72}
$q_0 = {\bf v}{\bf q} \equiv v q_L$
(following from the equality $qP=0$), strongly correlating
the energy difference $q_0$ with the longitudinal momentum
difference $q_L$,
the correlation function at $v\tau_0 > r_0$ substantially depends
on the direction of the vector  ${\bf q}$
even in the case of a
spherically symmetric spatial form of the production region.

Note that the on-shell constraint makes the $q$-dependence
of the correlation function essentially three--dimensional
(particularly, in pair c.m.s., $q\Delta x=-2{\bf k}^*{\bf r}^*$)
and thus makes impossible the unique Fourier reconstruction
of the space--time characteristics of the emission process.
However, within realistic models,
the directional and velocity
dependence of the correlation function
can be used to determine both
the duration of the emission and the form
of the emission region \cite{KP72}, as well as - to reveal
the details of the
production dynamics (such as collective flows;
see, {\it e.g.}, \cite{PRA84,MAK87} and the reviews
\cite{WH99,cso02}).
For this, the correlation functions can be analyzed
in terms of the {out} (x), {side} (y) and {longitudinal} (z)
components of the relative momentum vector
${\bf q}=\{q_x,q_y,q_z\}$ \cite{pod83,osl-sys};
the {out} and {side} denote the transverse,
with respect to the reaction axis, components
of the vector ${\bf q}$, the {out} direction is
parallel to the transverse component of the pair three--momentum.
The corresponding correlation widths are
usually parameterized in terms
of the Gaussian correlation radii $R_i$,
\begin{equation}
{\cal R}(p_{1},p_{2})=
1+\lambda\exp(-R_x^2q_x^2-R_y^2q_y^2-R_z^2q_z^2
-R_{xz}^2q_xq_z)
\label{osl}
\end{equation}
and their dependence on pair rapidity and transverse momentum
is studied.
The form of Eq.~(\ref{osl}) assumes azimuthal symmetry of the
production process \cite{WH99,pod83}. Generally, {\it e.g.},
in case of the
correlation analysis with respect to the reaction plane, all three
cross terms $q_iq_j$ contribute \cite{wie_fi98}.

It is well known that particle correlations at high energies
usually measure only a small part of the space-time emission volume,
being only slightly sensitive to its increase related to the fast
longitudinal motion of particle sources. In fact,
due to limited source decay momenta
${\rm p}^{(s)}$
of few hundred MeV/c, the correlated particles with nearby velocities
are emitted by almost comoving sources and so - at nearby space--time
points.
In other words, the maximal contribution of the relative motion
to the correlation radii in the two--particle c.m.s. is limited
by the moderate source decay length $\tau {\rm p}^{(s)}/m$.
The dynamical examples are sources-resonances,
colour strings or hydrodynamic expansion.
To substantially eliminate the effect of the longitudinal motion,
the correlations can be
analyzed in terms of the invariant variable
$q_{inv}\equiv Q = (-\widetilde{{q}}^2)^{1/2} = 2k^*$ and
the components of the  momentum difference in pair c.m.s.
(${\bf q}^*\equiv {\bf Q}= 2{\bf k}^*$) or in
the longitudinally comoving system (LCMS) \cite{cso91}.
In LCMS each pair is emitted transverse to the reaction axis
so that the generalized relative momentum
$\widetilde{{\bf q}}$ coincides with ${\bf q}^*$
except for the component
$\widetilde{q}_x=\gamma_{t}q_x^*$,
where $\gamma_{t}$ is the LCMS Lorentz factor of the pair.

Particularly, in the case of one--dimensional boost invariant
expansion, the longitudinal correlation radius in the LCMS reads
\cite{MAK87} $R_z \approx (T/m_t)^{1/2}\tau$, where $T$ is the
freeze-out temperature, $\tau$ is the proper freeze-out time and
$m_t$ is the transverse particle mass.
In this model, the side radius measures the
transverse radius of the system while,
similar to Eq.~(\ref{5}), the square of the out radius
gets an additional contribution $(p_t/m_t)^2\Delta\tau^2$
due to the finite emission duration $\Delta\tau$.
The additional transverse expansion leads to a slight
modification of the $p_t$--dependence of the longitudinal radius and -
to a noticeable decrease of the side radius and the spatial
part of the out radius with $p_t$.
Since the freeze-out temperature and the transverse flow determine
also the shapes of the $m_t$-spectra,
the simultaneous analysis of correlations and single particle
spectra for various particle species allows to disentangle
all the freeze-out characteristics
(see the review \cite{WH99}). It appears that with the increasing
energy of heavy ion collisions from AGS and SPS
up to the highest energies at RHIC,
the data show rather weak energy dependence \cite{adl01}
and point to the kinetic
freeze-out temperature somewhat below the pion mass,
a strong transverse flow (with the mean transverse flow velocity
at RHIC exceeding half the velocity of light \cite{nxu02}),
a short evolution time of 8-10 fm/c and a very short
emission duration of about 2 fm/c.
The short evolution and emission duration at RHIC are also supported
by the correlation analysis with respect to the reaction plane
\cite{lis03}.
The small time scales at RHIC were not expected in
transport and hydrodynamic models \cite{sbd01,hk02} and may
indicate an explosive character of particle production
(see, {\it e.g.}, \cite{cso94,dum02}).
In fact, the RHIC data can be described in so called blast wave
model \cite{adlr01,ret01} assuming a strong three-dimensional expansion
with a sharp boundary of the freeze-out density profile
in transverse plane. The same model with $\sim 15\%$ lower
mean transverse flow velocity is also consistent with
the SPS data \cite{twh99}.

\section{Multiboson and coherence effects}

In present and future heavy ion experiments at SPS, RHIC and LHC
many hundreds or thousands of pions can be produced per a unit
rapidity interval.
Since pions are bosons
there can be multiboson effects
enhancing the production of pions with low relative momenta
thus increasing the pion multiplicities, softening their spectra and
modifying the correlation functions
(see \cite{pra93,mb00,hei00} and references therein).
In particular,
it was shown \cite{mb00} that the width of the low-$p_t$ enhancement
due to BE condensation decreases with the system size as
$r_0^{-1/2}$ and this narrowing makes easier the identification
of this effect among others.
For the events of approximately fixed multiplicity,
the multiboson effects
can be triggered by decreasing correlation strength and a dip in the
two--pion correlation function at intermediate relative momenta
\cite{mb00,hei00}.

Though the present data does not point to any spectacular
multiboson effects, one can hope to observe new interesting
phenomena like boson condensation or speckles in some rare
events or in eventually overpopulated kinematic regions with
the pion density in the 6-dimensional phase space,
$f=(2\pi)^3 d^6n/d^3{\bf p}d^3{\bf x}$,
of the order of unity.
An example is a rapidly expanding system with the entropy much
smaller than in the case of total equilibrium.
Then a strong transverse flow can lead to rather dense gas of
soft pions in the central part of the
hydrodynamic tube at the final expansion stage
(see, {\it e.g.}, \cite{akk95}).
Another reason can be the expected formation of quark-gluon
plasma or mixed phase.
Due to large gradients of temperature
or velocity the hydrodynamic layer near the boundary with vacuum
can decay at a large phase space density and lead to pion speckles
even at moderate transverse momenta \cite{sla91}.

In the low-density limit ($f\ll 1$), the mean phase space density
at a given momentum ${\bf p}$ can be estimated as the mean number
of pions interfering with a pion of momentum ${\bf p}$
(rapidity $y$ and transverse momentum ${\bf p}_t$)
and building the Bose-Einstein (BE)
enhancement in the two-pion correlation function
\cite{ber94,sl94}:
$\langle f\rangle _{{\bf p}}\sim \pi^{3/2}{N}({\bf p})/V$,
where ${N}({\bf p})=d^3{n}/d^3{\bf p}$ and
$V=r_xr_yr_z$ is the interference volume defined in terms of the
outward ($r_x$), sideward ($r_y$) and longitudinal ($r_z$)
interferometry radii.
Typically $\langle f\rangle _{{\bf p}}\sim 0.1$
for mid-rapidities and $p_t\sim \langle p_t\rangle$ \cite{ber94}.
The data are also consistent with the phase space density of pions
near the local thermal equilibrium \cite{bar97,fer99}.

At AGS and SPS energies the interference volume $V$ seems to
scale with $dn/dy$ (see, {\it e.g.}, \cite{fer99,gaz95})
pointing to the freeze-out of pions at a constant phase space density.
This trend is however questioned by recent STAR data from RHIC,
indicating an increase of the freeze-out phase space density
with energy (a slight increase of $V$ is
not sufficient to balance $\sim 50 \%$ increase of $dn/dy$
as compared with SPS) and centrality \cite{ray02}.
Extrapolation of the RHIC phase space density measurements to
low transverse momenta predicts $\langle f\rangle _{{\bf p}}$
close to unity for central events, suggesting that significant
multiboson effects can be present at low $p_t$ at RHIC.

According to lattice Monte Carlo calculations including dynamical
fermions, deconfining phase transition leading to a 
quark-gluon plasma (QGP) phase of
matter is accompanied by restoration of chiral symmetry.
Subsequent phase transition into the hadronic phase can be revealed,
particularly, through substantial delays in particle emission and/or,
through the coherent component of the pion radiation.
This component would be characterized by a narrow Poisson
multiplicity distribution, contrary to wide multiplicity fluctuations
in the usual BE condensate.
The pions in the coherent state may
appear from the decay of a quasi-classical pion field (the order
parameter of the phase transition), the latter possibly related
to the spontaneous chiral symmetry breaking via the formation of
the disoriented chiral condensate (DCC) (see \cite{bjo97} and
a review \cite{bk96}).

The most plausible mechanism of DCC formation is a fast expansion
of hot QGP
resulting in a rapid supression of thermal fluctuations (quenching),
which in turn triggers a dramatic amplification of soft pion modes.
The detection and study of DCC is expected to provide valuable
information about the chiral phase transition and vacuum structure
of strong interactions.
DCC formation is usually expected to be associated with large
event-by-event fluctuations in the ratio of neutral to charged pions
in a certain phase--space domain.  The search for these fluctuations
at CERN SPS has so far resulted in setting only an
upper limit  on the production of a {\it single} DCC domain
\cite{agg98}. The absence of experimental evidence
for isospin fluctuations has been
however recently claimed to be in agreement with presumably more
realistic picture of an "unpolarized" DCC with the Fourier
modes of the field randomly oriented in isospin space
(instead of being aligned as in the original DCC scheme) \cite{ser03}.
The search for other DCC signatures like low momentum
pion clusters is therefore important.
Particularly, one can exploit the impact of the admixture of
coherent radiation on the QS and Coulomb correlations of like and
unlike pions \cite{als02}.
Other possibilities of experimental investigations of BE condensate
and DCC phenomena have been discussed, {\it e.g.},
in \cite{akk99,bz99b}.

The presence of the coherent pions (or pions emitted in the same
quantum state) manifests itself also
as a suppression of the BE correlations of two or more identical
pions \cite{gkw79,fw77,llp83,lyu91}.
Unfortunately, there are also other reasons leading to the
suppression of particle correlations. Besides the experimental
effects like finite resolution and particle misidentification
(that can be corrected for), presumably the most important one
is the contribution of the
particles emitted by long-lived sources \cite{lp79},
leading to the appearance of the parameter $\lambda < 1$ in
Eqs. (\ref{5}) and (\ref{osl}).
Also the usual Gaussian parameterizations of the
QS correlation functions may be inadequate and
lead to $\lambda < 1$ in the presence of the sources
with moderate but very different space-time characteristics
\cite{lp79,lp92,clz96}.

In principle, the effect of long-lived sources can be eliminated
in a combined analysis of two--pion and three--pion correlation
functions. The measured quantity is the genuine three--pion
correlation normalized with the help of the three two--pion
contributions - its intercept measures the chaotic or
coherent fraction \cite{hz97}.
First such measurements have been done only recently
in heavy ion experiments at CERN SPS \cite{bea99,agg00}
and RHIC \cite{wil02} and, in $e^+e^-$ collisions at LEP \cite{L3}.
The most accurate ones at RHIC and LEP indicate a dominant
chaotic fraction though the systematic errors allow for a substantial
coherent component.
Some sources of the systematic errors,
{\it e.g.}, the simplified treatment of the two-body
Coulomb and strong FSI,
can be overcome. However others, {\it e.g.}, the approximate
(factorization) treatment of the multiparticle FSI or
the insufficiently differential analysis of the three-pion
correlation function, can hardly be avoided at present computational
and experimental possibilities.

\section{Femtoscopy with unlike particles}

The complicated dynamics of particle production,
including resonance decays and particle rescatterings,
leads to essentially non--Gaussian tail of the
distribution of the relative distances $r^*$
of the particle emitters in the pair rest frame.
Therefore, due to different $r^*$--sensitivity
of the QS, strong and Coulomb FSI effects,
one has to be careful when analyzing the correlation functions
in terms of simple models.
Thus, the QS and strong FSI effects are influenced by the $r^*$--tail
mainly through the suppression parameter $\lambda$
already for distances of the order of inverse $q$-resolution
(typically some tens fm)
while, the Coulomb FSI is sensitive to the distances as large as
the pair Bohr radius $|a|$;
for $\pi\pi$, $\pi K$, $\pi p$, $KK$, $K p$ and $p p$ pairs,
$|a|=$ 387.5, 248.6, 222.5, 109.6, 83.6 and 57.6 fm,
respectively.
Clearly, the usual Gaussian parameterizations
of the distributions of the components
of the distance vector ${\bf r}^*$ may lead to
inconsistencies in the treatment of QS and FSI effects
(the Coulomb FSI contribution requiring larger effective radii).
These problems can be at least partially overcome with the help of
transport code simulations accounting for the dynamical evolution
of the emission process and providing the phase space information
required to calculate the QS and FSI effects on the correlation
function.

Thus, in a preliminary analysis of the NA49
correlation data from
central $Pb+Pb$ 158 AGeV collisions \cite{lna49,led02},
the freeze--out phase space distribution has been simulated
with the RQMD v.2.3 code \cite{rqmd}. The correlation functions
have been calculated using the code of Ref. \cite{ll1},
weighting the simulated pairs by squares of the
corresponding wave functions.
The dependence of the correlation function on the invariant
relative momentum $Q=2k^*$
was than fitted according to the formula \cite{lna49}
\begin{equation}
{\cal R}(Q)= {\rm norm}\cdot [{\rm purity}\cdot
{\rm RQMD}(r^*\rightarrow {\rm scale}\cdot r^*)+ (1-{\rm
purity})];
\label{Q}
\end{equation}
to account for a possible mismatch in $\langle r^*\rangle$,
the dependence on the $r^*$--scale parameter has been
introduced using the quadratic interpolation of the points simulated
at three scales chosen at 0.7, 0.8 and 1.
The fitted values of the purity parameter are in reasonable agreement
with the expected contamination of $\sim 15\%$
from strange particle decays and particle misidentification.
The fitted values of the scale parameter indicate that RQMD
overestimates the distances $r^*$ by 10-20$\%$. Similar
overestimation has been also observed when comparing RQMD
predictions with the NA49 data on $pp$ and $\pi^\pm\pi^\pm$
correlations \cite{ppna49,gan99,pipina49}.

Recently, there appeared data on $p\Lambda$ correlation functions
from $Au+Au$ experiment E985 at AGS \cite{lis01} and
$Pb+Pb$ experiment NA49 at SPS CERN \cite{blu02}.
As the Coulomb FSI is absent
in $p\Lambda$ system, one avoids here the problem of its sensitivity
to the $r^*$--tail. Also, the absence of the Coulomb suppression
of small relative momenta makes this system more sensitive to the
radius parameters as compared with $pp$ correlations \cite{wan99}.
In spite of rather large statistical errors,
a significant enhancement is seen at low relative momentum,
consistent with the known singlet and triplet
$p\Lambda$ s--wave scattering lengths.
In fact, the fits using the analytical expression for the
correlation function (originally derived for $pn$ system
\cite{ll1}) yield for the AGS data \cite{led02} the
purity of $0.5\pm 0.2$ and
the Gaussian radius of $4.5 \pm 0.7$ fm. For the NA49
data the fitted parameters are \cite{blu02} $0.17\pm 0.11$
and $2.9 \pm 0.7$ fm.
The fitted AGS purity is
consistent with the estimated one,
while the NA49 purity is about one standard
deviation too low. Fixing the NA49 purity at the estimated value
of 0.33, the Gaussian radius increases
by about 1 fm and becomes $3.8 \pm 0.4$ fm \cite{blu02}.
The fitted AGS and NA49 radii are in agreement with the radii
of 3-4 fm obtained from $pp$ correlations in
heavy ion collisions at GSI, AGS and SPS energies.

\section{Correlation measurement of strong interaction}
In case of a poor knowledge of the two--particle strong interaction,
which is the case for meson--meson, meson--hyperon or
hyperon--hyperon systems,\footnote
{
The $\Lambda\Lambda$ system is of particular interest
in view of an experimental
indication on the enhanced $\Lambda\Lambda$ production near
threshold \cite{ahn98}
and its possible connection with  the 6-quark H dibaryon problem.
}
it can be improved with the help of
correlation measurements.

In heavy ion collisions, the effective radius $r_0$
of the emission region can be
considered much larger than the range of the
strong interaction potential.
The FSI contribution is then independent of the actual
potential form \cite{gkll86}. At small $Q=2 k^*$,
it is determined by
the s-wave scattering amplitudes $f^S(k^*)$
\cite{ll1}.
In case of $|f^S|>r_0$, this contribution is of the order of
$|f^S/r_0|^2$ and dominates over the effect of QS.
In the opposite case,
the sensitivity of the correlation function to the scattering
amplitude is determined by the linear term $f^S/r_0$.

The possibility of the correlation measurement
of the scattering amplitudes has been demonstrated \cite{led02}
in a recent analysis of the
NA49 $\pi^+\pi^-$ correlation data within the RQMD model.
For this, the strong interaction scale 
has been introduced (similar to the $r^*$-scale),
redefining the original s-wave $\pi^+\pi^-$scattering length
$f_0=$ 0.232 fm:
$f_0\rightarrow {\rm sisca}\cdot f_0$.
The fitted parameter sisca $=0.63\pm 0.08$ appears to be
significantly lower than unity. To a similar shift ($\sim 20\%$)
point also the recent BNL data on $K_{l4}$ decays \cite{pis01}.
These results are in agreement with the two--loop calculation
in the chiral perturbation theory with a standard value of the
quark condensate \cite{col00}.

Recently, also the singlet
$\Lambda\Lambda$ s--wave scattering length $f_0$ has been
estimated \cite{led02,blu02} based on the fits
of the NA49 $\Lambda\Lambda$ data.
Using the analytical expression for the correlation
function \cite{led99}
(originally derived for $nn$ system \cite{ll1})
and fixing the purity of direct
$\Lambda$--pairs at the estimated value of 0.16 and varying
the effective radius $r_0$ in the acceptable range of several fm,
one gets \cite{blu02}
{\it e.g.}, $f_0=2.4\pm 2.1$ and $3.2\pm 5.7$ fm
for $r_0=2$ and 4 fm respectively
(we use the same sign convention as for
meson--meson and meson--baryon systems).
Though the fit results are not very restrictive, they
likely exclude
the possibility of a large positive singlet scattering length
comparable to that of $\sim$20 fm for the two--nucleon system.

The important information comes
also from $\Lambda\Lambda$ correlations at LEP
\cite{ale00}.
Here the effective radius $r_0$ is substantially smaller than the
range of the strong interaction potential, so the $\Lambda\Lambda$
correlation function is sensitive to the potential form
and requires the account of the waves with orbital angular momentum
up to $l\sim 20$ \cite{lqm02}.
In Ref. \cite{ale00}, the strong interaction has been neglected
and the observed decrease of the $\Lambda\Lambda$ correlation
function at small $Q$
has been attributed solely to the effect of the QS
(Fermi-Dirac) suppression.
The correlation function has been fitted by the expression\footnote
{
The singlet and triplet contributions to the correlation
function ${\cal R}={\cal R}_s+{\cal R}_t$ are
${\cal R}_{s,t}=\widetilde{\rho}_{s,t}[1\pm\lambda\exp(-r_0^2Q^2)]$,
where $\widetilde{\rho}_{s,t}$ depend on the $\Lambda$-polarization
$\mbox{\boldmath${\cal P}$}$ according to Eq.~(\ref{pol}) with
$\mbox{\boldmath${\cal P}$}_1=\mbox{\boldmath${\cal P}$}_2=
\mbox{\boldmath${\cal P}$}$.
}
\begin{equation}
\label{l3fit}
{\cal R}=
1-\frac12 \lambda (1+\mbox{\boldmath${\cal P}$}^2)
\exp(-r_0^2Q^2)
\end{equation}
corresponding to the simple Gaussian distribution of the
components of the relative distance vector ${\bf r}^*$
characterized by a dispersion $2r_0^2$.
The fit results are however
unsatisfactory for two reasons \cite{lqm02}:
(i) the parameter $\lambda=1.2\pm 0.2$ (neglecting in
Eq.~(\ref{l3fit})
the $\mbox{\boldmath${\cal P}$}^2$ polarization term
on a percent level)
is significantly higher than the value of $\sim 0.5$ expected
due to the feed--down from $\Sigma^0$ and weak decays;
(ii) the parameter $r_0=0.11\pm 0.02$
fm appears to be smaller than the string model lower limit
of $\sim 0.2$ fm.
Therefore, the observed anti-correlation at small $Q$
can be considered as a direct evidence for a repulsive
core in the $\Lambda\Lambda$ interaction potential.\footnote
{
The repulsive core arises due to the exchange of vector mesons
and is present,
{\it e.g.}, in various Nijmegen potentials used for the analysis
of the double $\Lambda$ hypernuclei. The core height and width
are about 9 GeV and 0.4 fm respectively. The s--wave scattering length
(effective radius) ranges from about 0.3 (15) fm to 11 (2) fm.
}
In fact, reasonable fits can be achieved using the Nijmegen singlet
potential NSC97e \cite{fg02}, rescaling the triplet one from
Ref. \cite{hiy} and, neglecting spin-orbit and tensor couplings.
For example, at a fixed $\lambda=0.6$, the fitted radius takes
an acceptable value
$r_0=0.29\pm 0.03$ fm \cite{lqm02}.

\section{Accessing relative space--time asymmetries}

The correlation function of two non--identical particles,
compared with the identical ones,
contains a principally new piece of information on the relative
space-time asymmetries in particle emission
such as mean relative time delays in the emission
of various particle species \cite{LLEN95}.
It can be particularly useful in searches for the effects
of the quark-gluon plasma phase transition like delays
between the emission of strange and antistrange particles
due to the process of strangeness distillation from the mixed phase.
The important information is contained also in the spatial part
of the asymmetry related, in particular, with the intensity
of the collective flow \cite{led02}.

Since the information on the relative space--time shifts
enters in the two--particle wave function through the terms odd in
${\bf k}^*{\bf r}^*\equiv {\bf p}_1^*({\bf r}_1^*-{\bf r}_2^*)$,
it can be accessed studying the correlation functions
${\cal R}_{+i}$ and ${\cal R}_{-i}$
with respectively positive and negative projection $k^*_i$
of the momentum ${\bf k}^*={\bf p}_1^*=-{\bf p}_2^*$
on a given direction ${\bf i}$ or, - the
ratio ${\cal R}_{+i}/{\cal R}_{-i}$.
For example, ${\bf i}$ can be the direction of the pair
velocity or, any of the out (x), side (y), longitudinal (z)
directions. Note that in the LCMS system,
\begin{equation}
\label{51a}
r_x^*\equiv\Delta x^*=\gamma_{t}(\Delta x-v_{t}\Delta t),~~
r_y^*\equiv\Delta y^*=\Delta y,~~r_z^*\equiv\Delta z^*=\Delta z,
\end{equation}
where $\gamma_{t}=(1-v_t{}^2)^{1/2}$ and $v_t=P_t/P_0$
are the pair LCMS Lorentz factor and velocity.
One may see that the asymmetry in the out (x) direction
depends on both space and time asymmetries
$\langle\Delta x\rangle$ and $\langle\Delta t\rangle$.
In case of a dominant Coulomb FSI, the intercept of the correlation
function ratio is directly related with the asymmetry
$\langle r^*_i\rangle$ \cite{lpx,vol97} (see also \cite{alinote95}):
\begin{equation}
{\cal R}_{+i}/{\cal R}_{-i}\approx 1+
2\langle r_i^*\rangle /a,
\label{51}
\end{equation}
where $a=(\mu z_{1}z_{2}e^{2})^{-1}$ is the Bohr radius of the
two-particle system taking into account the sign of the interaction
($z_ie$ are the particle electric charges,
$\mu$ is their reduced mass).

At low energies, the particles in heavy ion collisions are emitted
with the characteristic emission times of tens to hundreds fm/c so
that the observable time shifts should be of the same order
\cite{LLEN95}.
Such shifts have been indeed observed with the help of the
${\cal R}_{+}/{\cal R}_{-}$ correlation ratios for
proton-deuteron systems in several heavy ion
experiments at GANIL \cite{ghi95} indicating,
in agreement with the coalescence model, that deuterons are
on average emitted earlier than protons.

For ultra-relativistic heavy ion collisions,
the sensitivity of the ${\cal R}_+/{\cal R}_-$ correlation ratio to
the relative time shift
$\langle\Delta t\rangle$ (introduced {\it ad hoc})
was studied for various two-particle systems simulated using the
transport codes \cite{alinote95}.
The scaling of the effect with the space-time asymmetry and
with the inverse Bohr radius $a$ was clearly illustrated.
It was concluded that
the ${\cal R}_{+}/{\cal R}_{-}$ ratio can be sensitive to the shifts
in the particle emission times of the order of a few fm/c.
Motivated by this result, the correlation asymmetry
for the $K^+K^-$ system has been studied
in a two-phase thermodynamic evolution model and
the sensitivity has been demonstrated
to the production of the transient strange quark matter state
even if it decays on strong interaction time scales \cite{sof97}.
The method sensitivity to the space-time asymmetries arising
also in the
usual multiparticle production scenarios was demonstrated
for AGS and SPS energies using the transport code RQMD
\cite{lna49,lpx,vol97}.
At AGS energy, the $Au+Au$ collisions have been simulated and
the $\pi p$ correlations have been studied
in the projectile fragmentation region where proton directed
flow is most pronounced and where the proton and pion sources
are expected to be shifted
relative to each other both in the longitudinal
and in the transverse directions
in the reaction plane.
It was shown \cite{vol97} that
the corresponding ${\cal R}_{+}/{\cal R}_{-}$ ratios are
sufficiently sensitive to reveal the
shifts; they were confirmed in the directional analysis of
the experimental AGS correlation data \cite{mis98}.

At SPS energy, the simulated central $Pb+Pb$ collisions
yield practically zero asymmetries for $\pi^+\pi^-$ system
while, for $\pi^\pm p$ systems, the LCMS asymmetries
are
$\langle\Delta x\rangle = -6.2$ fm,
$\langle\Delta y\rangle =\langle\Delta z\rangle =0$,
$\langle\Delta t\rangle = -0.5$ fm/c,
$\langle\Delta x^*\rangle = -7.9$ fm in the symmetric
midrapidity window\footnote
{
$\langle\Delta y\rangle=0$ due to
the azimuthal symmetry and
$\langle\Delta z\rangle=0$ in a symmetric mid--rapidity window
due to the symmetry of the initial system.
}
\cite{lpx} and,
$\langle\Delta x\rangle = -5.2$ fm,
$\langle\Delta y\rangle =0$,
$\langle\Delta z\rangle = -6.5$ fm,
$\langle\Delta t\rangle = 2.9$ fm/c,
$\langle\Delta x^*\rangle = -8.5$,
for the NA49 acceptance (shifting the rapidities into the
forward hemisphere) \cite{lna49}.
Besides, $\langle x\rangle$ increases with particle $p_t$ or
$u_t=p_t/m$, starting from zero due to kinematic reasons.
The asymmetry arises because of a faster increase with $u_t$
for heavier particle.
The non--zero positive value of
$\langle x\rangle=\langle {\bf r}_t\hat{\bf x}\rangle$
($\hat{\bf x}={\bf p}_t/p_t$ and ${\bf r}_t$ is the transverse
radius vector of the emitter) and the hierarchy
$\langle x_\pi\rangle<\langle x_K\rangle<\langle x_p\rangle$
is a signal of a universal transversal collective flow
\cite{lna49,led02}.
To see this, one should simply take into account that the
thermal transverse
velocity $\beta_T$ is smaller for heavier particle
and thus washes out the positive shift due to the transversal
collective flow velocity $\beta_F$ to a lesser extent.
More explicitly, in the non--relativistic approximation,
the transverse velocity
$\mbox{\boldmath$\beta$}_t\doteq \mbox{\boldmath$\beta$}_F+
\mbox{\boldmath$\beta$}_T$; in the out-side decomposition,
$\mbox{\boldmath$\beta$}_t=\beta_t \{1,0\}$,
$\mbox{\boldmath$\beta$}_F=\beta_F \{\cos\phi_r,\sin\phi_r\}$,
$\mbox{\boldmath$\beta$}_T=\beta_T \{\cos\phi_T,\sin\phi_T\}$.
Due to the azimuthal symmetry,
the vector of the transversal collective flow velocity
$\mbox{\boldmath$\beta$}_F$ is parallel to the
transverse radius vector
${\bf r}_t=r_t \{\cos\phi_r,\sin\phi_r\}$ and, its magnitude
depends only on $r_t$:
$\beta_F=\beta_F(r_t)$.
To calculate $\langle x\rangle$, one has to average over four
variables $r_t$, $\phi_r$, $\beta_T$ and $\phi_T$.
At a fixed transverse velocity vector
$\mbox{\boldmath$\beta$}_t$,
only two of them ({\it e.g.}, $r_t,\phi_r$ or $r_t,\beta_T$)
are independent.
In particular,
$\beta_T^2=\beta_t^2+\beta_F^2-2\beta_t\beta_F\cos\phi_r,$
so the destructive effect of the thermal velocity
$\beta_T$
on the out shift is clearly seen:
\begin{equation}
\label{51b}
\langle x\rangle=
\langle r_t\cos\phi_r\rangle=
\left\langle r_t\frac{\beta_t^2+\beta_F^2-\beta_T^2}
{2\beta_t\beta_F}\right\rangle.
\end{equation}
The maximal out shift $\langle x\rangle_{\max}=\langle r_t\rangle$
corresponds to zero thermal velocity. The shift vanishes when
the width of the contributing interval
$|\beta_t-\beta_F|\le \beta_T\le\beta_t+\beta_F$
becomes negligible compared with the characteristic width
of the thermal distribution, {\it e.g.},
at $\beta_t\rightarrow 0$ or $\beta_F\rightarrow 0$
or, for very light particles; the angle $\phi_r$ is
then decorrelated from $\beta_T$ and so distributed
uniformly in the full angular interval $(-\pi,\pi)$.\footnote
{
Note that, irrespective of the thermal width,
the side shift $\langle y\rangle=
\langle r_t\sin\phi_r\rangle=0$ since, due to azimuthal symmetry,
the angles $\phi_r$ and $-\phi_r$ contribute with the same
weights.
}
As a result, in case of a locally equilibrated expansion process,
one expects a negative asymmetry $\langle\Delta x\rangle\equiv
\langle x_1-x_2\rangle$ provided $m_1<m_2$. Moreover, this asymmetry
vanishes in both limiting cases: $\beta_F\ll \beta_T$ and
$\beta_F\gg \beta_T$.

These conclusions agree with the calculations
in the longitudinal-boost invariant
hydrodynamic model. Thus, assuming a linear non-relativistic
transversal flow velocity profile
$\beta_F=\beta_0 r_t/r_0$, the local thermal momentum distribution
characterized by the kinetic freeze-out temperature $T$
and the Gaussian density profile
$\exp(-r_t^2/(2r_0^2))$,
one confirms a faster rise of $\langle x\rangle$ with
$\beta_t$ for heavier particles
(see the non-relativistic limit of Eq.~(30) in Ref.~\cite{akk96}):
\begin{equation}
\label{x_hydro}
\langle x\rangle=
r_0\frac{\beta_t\beta_0}{\beta_0^2+T/m_t}.
\end{equation}
The maximal magnitude of the asymmetry
$\langle x_1-x_2\rangle$ at $\beta_{1t}=\beta_{2t}=v_t$
is achieved for an optimal value of the flow parameter
$\beta_0=T/(m_{1t}m_{2t})^{1/2}=T/(\gamma_t^2m_{1}m_{2})^{1/2}$;
{\it e.g.}, for $\pi p$ pairs at $v_t=0.6$
(close to a mean LCMS velocity of low-$Q$ $\pi p$ pairs
in the NA49 experiment at SPS \cite{lna49})
and $T= 120$ MeV,
the optimal value $\beta_0= 0.27$.
The SPS data on particle spectra and interferometry
radii in central $Pb+Pb$ collisions at 158 AGeV
are consistent with the parameters
$\beta_0\approx 0.35$, $r_0\approx 6$ fm and
$T\approx 120$ MeV with the
uncertainties of $10-20\%$ \cite{WH99,twh99,abs02}.
The corresponding out asymmetry for $\pi p$ pairs
$\langle \Delta x\rangle=
\langle x_\pi-x_p\rangle\approx -4$ fm
at $v_t=0.6$.
As for the longitudinal and time shifts,
in the longitudinal-boost invariant hydrodynamic model
$z=\tau\sinh\eta$ and $t=\tau\cosh\eta$, where $\tau$
is the proper freeze-out time and $\eta$ is the emitter
rapidity. At a given $p_t$, the LCMS $\eta$-distribution of the
contributing emitters is given by the thermal law
$\exp(-m_t\cosh\eta/T)$. Being symmetric, it predicts
vanishing longitudinal shift:
$\langle z\rangle=\langle \tau\sinh\eta\rangle=0$.
To estimate the time shift, for $m_t>T$ one can write
$\cosh\eta\approx 1+\eta^2/2$ and get
$\langle t\rangle\approx \tau(1+\frac12T/m_t)$.\footnote
{
One also recovers the expression for the LCMS
interferometry longitudinal radius squared  \cite{MAK87}:
$R_z^2=\langle (z-\langle z\rangle)^2\rangle\approx\tau^2 T/m_t$
up to a relative correction ${\cal O}(T/m_t)$.
}
For the central $Pb+Pb$ collisions at SPS, $\tau \sim 8$ fm/c
and the relative
time shift $\langle \Delta t\rangle=
\langle t_\pi-t_p\rangle\approx 3$ fm/c.
This shift is about the same as predicted by RQMD for the
asymmetric NA49 rapidity acceptance.
The magnitude of the relative out shift in pair rest frame
(determining the observable asymmetry),
$\langle \Delta x^*\rangle\approx -7$ fm, is however lower
than in RQMD due to $\sim 20\%$ lower magnitude
of $\langle \Delta x\rangle$.

In fact, the NA49 data on ${\cal R}_{+x}/{\cal R}_{-x}$ ratio
for $\pi^+p$ and $\pi^-p$ systems show consistent mirror symmetric
deviations from unity, their size of several percent and the
$Q$--dependence being in agreement with RQMD calculations corrected
for the resolution and purity
\cite{led02,pipina49,blu02}.
Similar pattern of the correlation asymmetries has been reported
also for $\pi^\pm K^\pm$ and $\pi^\pm K^\mp$ systems
in experiment STAR at RHIC.
They seem to be in agreement with the hydrodynamic type calculations
with a stronger transverse flow than at SPS
and a box-like density profile (blast wave), and -
somewhat lower than RQMD predictions
\cite{ret01,ray02}.

The finite widths of particle rapidity distributions
require however a violation of the boost invariance.
It can be parameterized by
a Gaussian dispersion $\Delta\eta^2$ of the LCMS
$\eta$-distribution centered at $-Y$, where
$Y$ is the CMS pair rapidity; {\it e.g.},
the data on central $Pb+Pb$ collisions at
158 AGeV are consistent with $\Delta\eta=1.3$ \cite{WH99}.
As a result,
\begin{equation}
\label{z_hydro}
\langle z\rangle\approx -\tau Y (1+\Delta\eta^2m_t/T)^{-1}
\end{equation}
and
$\langle t\rangle$ acquires a $Y$-dependent contribution
$\frac12 \tau Y^2(1+\Delta\eta^2 m_t/T)^{-2}$.
For the asymmetric NA49 rapidity acceptance,
the mean $\pi p$ pair rapidity $Y\sim 1.5$,
$\langle z_\pi-z_p\rangle\approx -2.8$ fm and
the $\pi p$ time shift at $Y=0$ is increased by $\sim 0.7$ fm/c.
This is in qualitative agreement with the
RQMD predictions for the rapidity dependence
of the longitudinal and time shifts.
The magnitude of the $Y$-dependent shifts in the
hydrodynamic model is however substantially smaller.
Besides, the LCMS emission times in RQMD are
by a factor of $2-3$ larger and show substantial dependence
on the transverse velocity \cite{lna49}.
These differences may point to the oversimplified space-time
evolution picture in the hydrodynamic model.
Particularly, the neglect of $r_t$-dependence of the proper
freeze-out time and of the longitudinal acceleration during
the evolution may not be justified \cite{WH99,akk96}.

\section{Spin correlations}

The information on the system size and the two--particle interaction
can be achieved also with the help of spin correlation measurements
using as a spin analyzer the asymmetric (weak) particle
decay \cite{lna49,ale95,led01}.
Since this technique requires no construction of the uncorrelated
reference sample, it can serve as an important consistency check
of the standard correlation measurements.
Particularly, for two $\Lambda$--particles decaying into the $p\pi^-$
channel characterized by the asymmetry parameter $\alpha=0.642$,
the distribution of the cosine of the relative angle $\theta$ between
the directions of the decay protons in the respective $\Lambda$ rest
frames allows one to determine the triplet fraction
$\rho_t={\cal R}_t/{\cal R}$,
where ${\mathcal R}_t$ is
the triplet part of the correlation function
(see the footnote in connection with Eq.~(\ref{l3fit})):
\begin{equation}
dN/d\cos\theta=\frac12\left[1+\alpha^2\left(\frac43\rho_t-1\right)
\cos\theta\right].
\end{equation}

Both the correlation and spin composition measurements were recently
done for two--$\Lambda$ systems produced in multihadronic
$Z^0$ decays at LEP \cite{ale00,LEP}.
Except for a suppression at  $Q < 2$ GeV/c,
the triplet fraction $\rho_t$ was
found to be consistent with the value 0.75,
as expected from a statistical spin mixture.
Such a suppression, as well as
similar suppression of the usual correlation function,
is expected due to the effects of QS and a repulsive
potential core, and points to a small correlation
radius $r_0 < 0.5$ fm \cite{lqm02}.

The spin correlations allow also for a relatively simple
test of the quantum--mechanical coherence, based on Bell--type
inequalities derived from the assumption of the factorizability
of the two--particle density matrix, {\it i.e.}
its reduction to a sum
of the direct products of one--particle density matrices with the
nonnegative coefficients \cite{led01}. Clearly, such a form of the
density matrix corresponds to a classical probabilistic description
and cannot account for the coherent quantum--mechanical effects,
particularly, for the production of two $\Lambda$-particles in a
singlet state. Thus
the suppression of the triplet $\Lambda\Lambda$ fraction observed
in multihadronic $Z^0$ decays at LEP
indicates a violation of one of the Bell-type inequalities
$\rho_t \ge 1/2$.

\section{Conclusions}

Thanks to the effects of quantum statistics and final state
interaction,
the particle momentum and, recently, also spin correlations
give unique information on the space--time
production characteristics and the collective phenomena like
multiboson and coherence effects and collective flows.
Besides the flow signals from single-particle spectra and
like-meson interferometry,
rather direct evidence for a strong transverse flow in heavy ion
collisions at SPS and RHIC comes from unlike particle
correlation asymmetries. Being sensitive to relative time delays
and collective flows,
the correlation asymmetries can be especially useful to study
the effects of the quark--gluon plasma phase transition.
The correlations yield also a valuable information on the
particle strong interaction hardly accessible by other means.


\end{document}